\documentclass{arxiv-nature}

\usepackage{graphicx}
\usepackage{amsmath}
\usepackage{microtype}
\usepackage{amssymb}

\providecommand{\aj}{Astronomical Journal}
\providecommand{\aap}{Astronomy \& Astrophysics}
\providecommand{\apj}{Astrophysical Journal}
\providecommand{\apjl}{Astrophysical Journal Letters}
\providecommand{\apjs}{Astrophysical Journal Supplement Series}
\providecommand{\araa}{Annual Review of Astronomy and Astrophysics}
\providecommand{\mnras}{Monthly Notices of the Royal Astronomical Society}
\providecommand{\nat}{Nature}
\providecommand{\prl}{Physical Review Letters}
\providecommand{\pasp}{Publications of the Astronomical Society of the Pacific}

\date{Submitted on 2026 September 18}

\begin{document}

\title{Stochastic black hole growth tracks stellar mass in ultramassive galaxies}

\author{Michele Cappellari$^1$\thanks{E-mail: michele.cappellari@physics.ox.ac.uk}, Dieu D. Nguyen$^2$\thanks{E-mail: dieun@umich.edu}, and Susan A. Kassin$^3$\\
$^1$Sub-Department of Astrophysics, Department of Physics, University of Oxford, Denys Wilkinson Building, Keble Road, Oxford, OX1 3RH, UK\\
$^2$Department of Astronomy, University of Michigan, 1085 South University Avenue, Ann Arbor, MI 48109, USA\\
$^3$Space Telescope Science Institute, 3700 San Martin Drive, Baltimore, MD 21218, USA}

\abstract{ 
Over the past three decades, it has been established that galaxy spheroids contain supermassive black holes (BHs) whose masses correlate tightly with the properties of their host galaxies. The most predictive of these correlations is between BH mass and stellar velocity dispersion (the $M_{\rm BH}-\sigma$ relation) \cite{Ferrarese2000, Gebhardt2000bh}, alongside a relation with bulge luminosity or stellar mass (the $M_{\rm BH}-M_\star$ relation) \cite{Marconi2003, Haring2004}. Theoretical models of galaxy evolution predict that for the most massive galaxies, whose late-time growth proceeds mainly via gas-poor `dry' mergers \cite{Oser2010,Naab2017}, $M_\star$ should become a better predictor of BH mass than $\sigma$ \cite{Lauer2007,Cappellari2011nat,Cappellari2013p20,Krajnovic2018channels}. However, genuine deviations from the $M_{\rm BH}-\sigma$ relation have remained difficult to observe due to the extreme rarity and low central surface brightness of such galaxies. Here we report James Webb Space Telescope (JWST) measurements of central BHs in a sample of eight ultramassive galaxies. We find the largest deviations from the $M_{\rm BH}-\sigma$ relation ever reported, with most BH masses exceeding $10^{10}$ solar masses. In this extreme mass regime, the galaxies tightly follow the $M_{\rm BH}-M_\star$ relation with minimal scatter, successfully confirming theoretical predictions. Furthermore, we observe significant stochasticity: For one ultramassive galaxy, the central dispersion drops rather than peaks, revealing no evidence for a central mass; while we can formally only place an upper limit, the data are entirely consistent with no BH at all. Crucially, this physical dichotomy is already evident directly in the raw kinematic maps prior to any fitting, and is robustly confirmed by our dynamical modelling. This unexpected diversity suggests that extreme multi-body dynamical processes, such as gravitational recoil or three-body ejections, significantly disrupt BH growth in the Universe's most massive galaxies.
}

\maketitle

\begin{figure*}
    \centering
    \includegraphics[width=\textwidth]{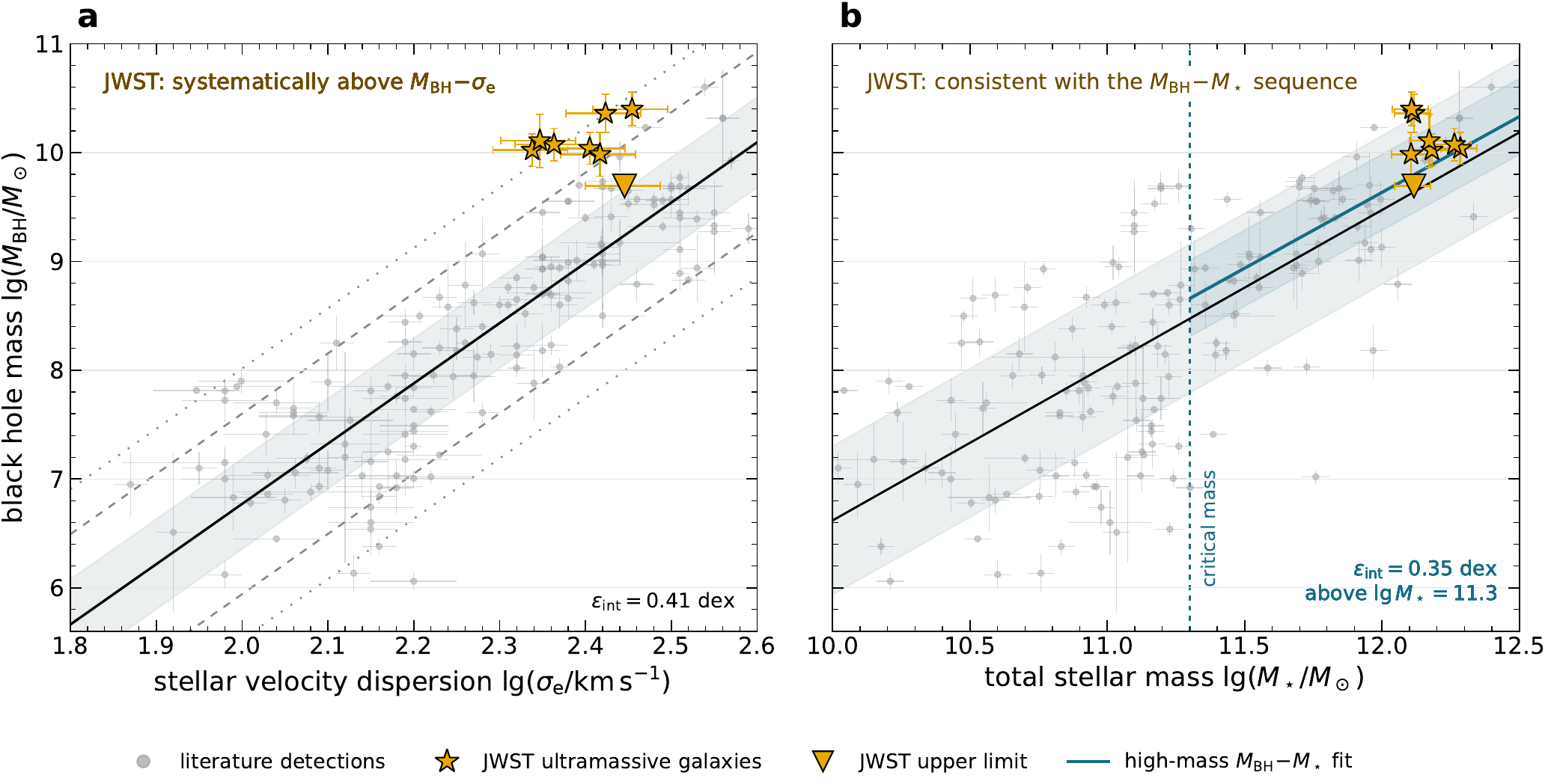}
    \caption{\textbf{A stochastic breakdown of the $M_{\rm BH}-\sigma_{\rm e}$ relation in ultramassive galaxies.} \textbf{a}, Black hole mass $M_{\rm BH}$ versus effective stellar velocity dispersion $\sigma_{\rm e}$. \textbf{b}, $M_{\rm BH}$ versus total stellar mass $M_\star$. Small grey symbols show literature detections, while large gold stars show the JWST ultramassive galaxies measured in this study. The gold downward triangle marks the JWST target for which the dynamical analysis yields an upper limit on $M_{\rm BH}$. In both panels, the black line and grey shaded region show the best-fitting relation and its $1\,\sigma$ intrinsic-scatter interval for the background comparison sample; grey dashed and dotted lines indicate $2\,\sigma$ and $3\,\sigma$ intervals. The JWST detections lie systematically above the $M_{\rm BH}-\sigma_{\rm e}$ relation, with several reaching offsets of $\sim3\,\epsilon_{\rm int}$. In panel \textbf{b}, the teal line and shading show the fit and intrinsic-scatter interval for galaxies with $\lg(M_\star/M_\odot)\geq11.3$, and the vertical teal line marks this transition mass. The JWST galaxies, including the upper limit, are consistent with the $M_{\rm BH}-M_\star$ sequence, showing that total stellar mass overtakes velocity dispersion as the primary predictor of black hole mass at the extreme high-mass end, while the upper limit demonstrates that this transition is stochastic rather than universal.}
    \label{fig:bh_relations}
\end{figure*}

The tight empirical scaling relations between supermassive black holes (BHs) and their host galaxies provide some of the strongest empirical evidence that the assembly of galaxies and the growth of their central BHs are intimately connected \cite{Kormendy1995, Magorrian1998}. The discovery of the $M_{\rm BH}-\sigma$ relation \cite{Ferrarese2000, Gebhardt2000bh}, which exhibits substantially smaller scatter than relations based on galaxy luminosity, historically suggested that the stellar velocity dispersion $\sigma$ is the most direct predictor of BH mass. However, theoretical models of hierarchical galaxy evolution indicate that this relationship should not remain universal across all mass scales. 

Ultramassive galaxies are particularly important in this context: as the end products of prolonged hierarchical assembly, they integrate the cumulative effects of successive growth events and provide unusually stringent constraints on models of galaxy evolution. For the most massive early-type galaxies, late-time evolution is expected to be dominated by dissipationless, or `dry', mergers rather than in-situ star formation \cite{Oser2010, Naab2017}. During a dry major merger, a simple virial argument ($\sigma^2 \propto M_\star / R_{\rm e}$) dictates that the resulting increase in effective radius ($R_{\rm e}$) largely compensates for the increase in stellar mass ($M_\star$), leaving the velocity dispersion relatively unchanged \citep{Hernquist1993,Ciotti2007,Naab2009,Bezanson2009}. Because the central BHs of the merging galaxies eventually coalesce, the final BH mass grows in direct proportion to the accumulated stellar mass. Consequently, repeated dry merging should move a galaxy to a larger $M_\star$ and $M_{\rm BH}$ at a nearly fixed $\sigma$, predicting a distinct bend in the $M_{\rm BH}-\sigma$ relation at the highest masses where $M_\star$ overtakes $\sigma$ as the better predictor of BH mass \cite{Lauer2007, Cappellari2011nat, Cappellari2013p20, Krajnovic2018channels}.

Detecting such an upturn at the high-mass end of the $M_{\rm BH}$--$\sigma_\star$ relation would constitute a landmark test of galaxy evolution, because it would directly probe whether the dominant channel of BH--galaxy growth changes during the late assembly of the most massive galaxies. Previous attempts have yielded intriguing but statistically limited evidence. Early studies identified hints of an upward departure in a few exceptionally luminous galaxies \cite{McConnell2011,McConnell2012,Thomas2016,Mehrgan2019}, while subsequent analyses used structural proxies for dry-merger histories, including depleted stellar cores \cite{Milosavljevic2001,Kormendy2004,Thomas2014} and slow rotation \cite{Emsellem2007,Emsellem2011p3}, to show that core-S\'ersic galaxies follow different $M_{\rm BH}$--$M_\star$ slopes from non-core galaxies \cite{Graham2012,Scott2013bh,Rusli2013cores,Dullo2021}. Major reviews of BH scaling relations have considered the possibility of an upturn in the high-mass $M_{\rm BH}$--$\sigma_{\rm e}$ relation but reached contrasting conclusions \cite{Kormendy2013review,Saglia2016,vandenBosch2016}. Even the latest synthesis finds only marginal empirical evidence for an upturn, despite predicting that a significant one should be observable, based on their core sizes \cite[fig.~2]{Dullo2026}. The rarity of ultramassive galaxies, combined with the low surface brightness of their expanded stellar cores, has made it extremely challenging for ground-based facilities to resolve the BH sphere of influence and obtain robust dynamical constraints on BH scaling relations at the top of the galaxy mass hierarchy \cite{Krajnovic2018channels}.

\begin{figure*}
    \centering
    \includegraphics[width=\textwidth]{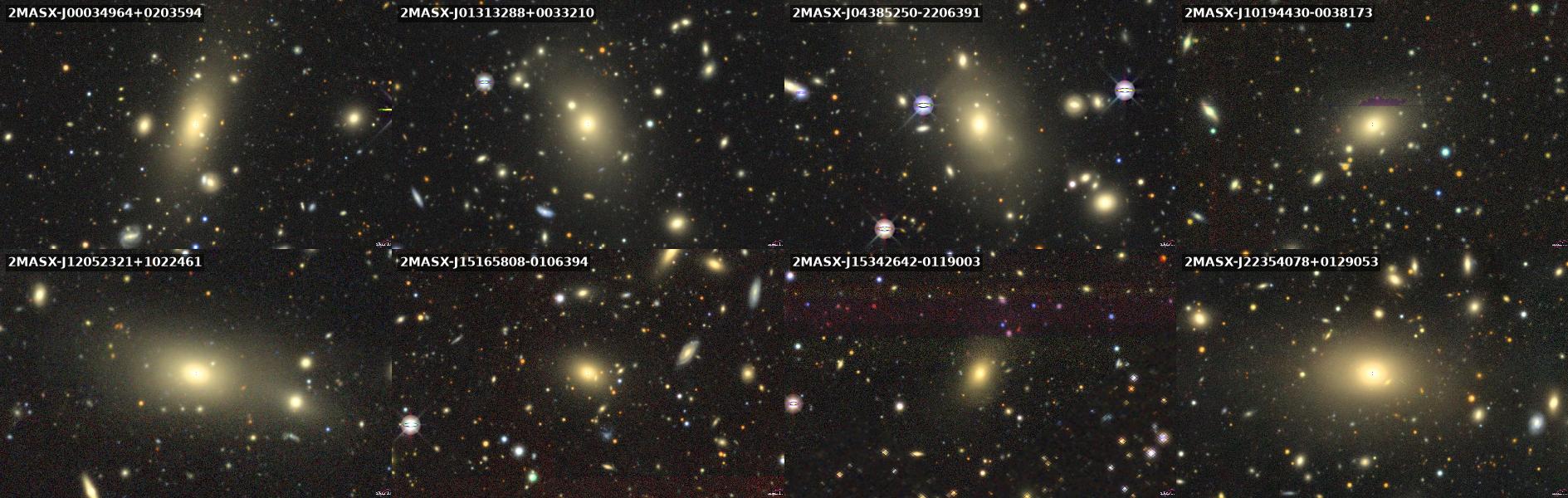} 
    \caption{\textbf{Large-scale environments of the JWST ultramassive galaxy sample.} Wide-field RGB composite images from the DESI Legacy Surveys (DR10) for the eight ultramassive galaxies in our sample. Each field of view spans approximately $13.8' \times 8.8'$, capturing environments that range from relative isolation to the dense cores of massive galaxy clusters. These deep, ground-based observations trace the extended stellar envelopes and intra-cluster light of these giant ellipticals, firmly anchoring the outer boundary conditions of our Multi-Gaussian Expansion (MGE) photometric models. Target identifiers are indicated in the top left of each panel. The images use standard DESI colour mapping, with the $g$, $r$ and $z$ bands assigned to the blue, green and red channels, respectively.}
    \label{fig:desi_rgb}
\end{figure*}

The launch of the James Webb Space Telescope (JWST) has opened a new horizon on this problem. By combining a large collecting area with exquisite, stable angular resolution, JWST allows us to resolve the gravitational signature of central BHs in distant galaxies, enabling us to sample sufficiently large cosmic volumes to discover these rare ultramassive systems. Here, we report direct dynamical BH mass measurements for a sample of eight ultramassive early-type galaxies, observed as part of the JWST Ultramassive Galaxy Sample, which was carefully selected from a parent sample or nearby ultramassive galaxies \cite{Nguyen2023}. These targets span a diverse range of large-scale environments, from relative isolation to the dense cores of massive galaxy clusters (Fig.~\ref{fig:desi_rgb}). These target galaxies are among the most massive objects for which BHs have ever been dynamically weighed; looking at the historical sample of roughly 150 reliable measurements, only three host galaxies surpass our targets in stellar mass.

Our homogeneous integral-field stellar kinematics reveal that the BH masses in these ultramassive galaxies represent the largest positive deviations from the canonical $M_{\rm BH}-\sigma$ relation ever reported (Fig.~\ref{fig:bh_relations}). Seven of the galaxies lie systematically and coherently above the extrapolation of the $M_{\rm BH}-\sigma$ relation, with four objects offset by roughly three times the intrinsic scatter ($\sim 3\,\epsilon_{\rm int}$) of the broad population. Consequently, their central BHs are among the most massive ever measured, with the majority exceeding $10^{10} M_\odot$. Prior to this JWST campaign, reaching this extreme mass threshold was exceptionally rare, with just three such ultramassive black holes confidently identified in the local universe.

Crucially, while these ultramassive galaxies dramatically depart from the $M_{\rm BH}-\sigma$ relation, they tightly trace the high-mass extension of the $M_{\rm BH}-M_\star$ relation. When we restrict the scaling relations above the critical mass $M_\star^{\rm crit} > 10^{11.3} M_\odot$ populated by slow-rotating, cored, dry-merger remnants \cite{Cappellari2016, Cappellari2026}, the $M_{\rm BH}-M_\star$ relation tightens considerably. In this regime, it exhibits an intrinsic scatter ($\epsilon_{\rm int} \approx 0.35$\,dex) that is smaller than that of the global $M_{\rm BH}-\sigma$ relation, and consistent to the tightness reported for the relation between BH mass and galaxy core size \cite{Dullo2026}. Our findings therefore provide the first definitive statistical evidence that $M_\star$ overtakes $\sigma$ as the superior predictor of BH mass at the extreme high-mass end. This transition strongly confirms theoretical predictions regarding the differential growth of BHs and galaxies once dry, dissipationless mergers become the dominant assembly channel.

\begin{figure*}
\centering
\includegraphics[width=\textwidth]{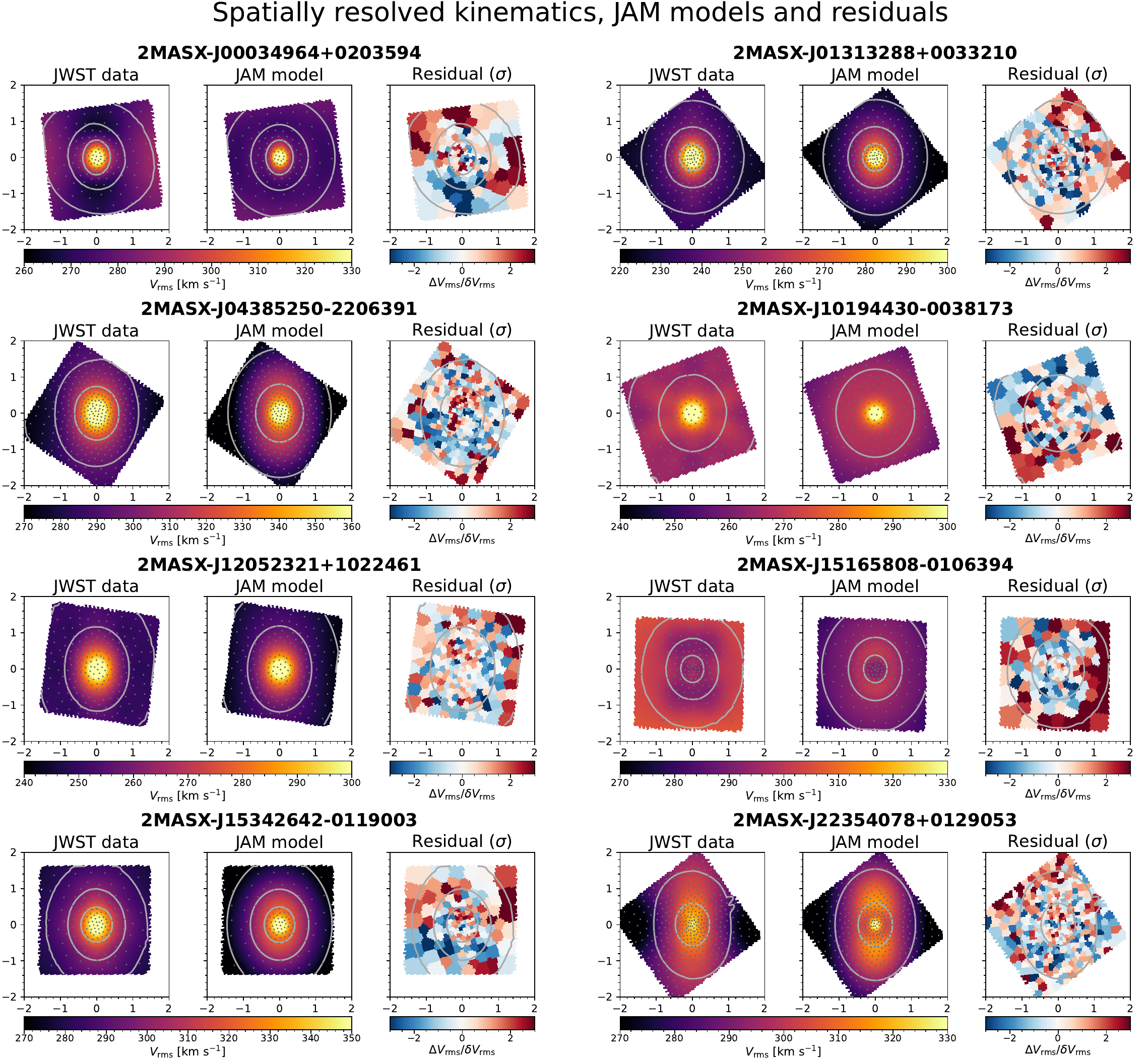}
\caption{\textbf{Spatially resolved stellar kinematics, JAM models, and residuals.} Comparison between the observed and JAM-modelled projected second velocity moment ($V_{\rm rms}=\sqrt{V^2+\sigma^2}$) for the eight ultramassive galaxies in our sample. For each galaxy (identifier above), three panels are shown: \textit{left:} even-symmetrized JWST/NIRSpec kinematic data extracted using BOSZ templates; \textit{middle:} best-fitting JAM model assuming prolate geometry and spherical alignment of the velocity ellipsoid \citep{Cappellari2026p2}; and \textit{right:} normalized residuals, defined as $\Delta V_{\rm rms}/\delta V_{\rm rms} = (V_{\rm rms}^{\rm data} - V_{\rm rms}^{\rm model})/\delta V_{\rm rms}$, evaluated directly at the original, non-symmetrized kinematic Power bins without interpolation or smoothing. Grey contours trace logarithmically spaced surface-brightness isophotes (in 1\,mag steps), and small grey dots mark the Power bin centroids. Coordinates are in arcseconds relative to the galaxy centre, rotated so that the galaxy photometric major/symmetry axis aligns with the $y$-axis. For each galaxy, the bottom-left colour bar indicates $V_{\rm rms}$ in $\mathrm{km\,s^{-1}}$ (common to both the data and model panels), while the bottom-right colour bar denotes the normalized residual.}
\label{fig:jam_maps_prolate}
\end{figure*}

However, the data also reveal that not all ultramassive galaxies host ultramassive BHs. Our sample displays a striking and genuine astrophysical diversity: while most targets contain vastly overmassive central objects, one displays no evidence for a central BH. Because completely ruling out a black hole is virtually impossible, we can formally place only an upper limit, consistent with a comparatively `normal' BH on the unextrapolated $M_{\rm BH}-\sigma$ relation, but entirely compatible with hosting no central BH at all. Crucially, this distinction is not merely a consequence of a loose statistical constraint: the kinematic data themselves appear qualitatively distinct and remarkably unusual. This dichotomy is directly visible in the spatially resolved kinematics and roughly captured by our best-fitting Jeans Anisotropic Models \cite[JAM][Fig.~\ref{fig:jam_maps_prolate}]{Cappellari2026jam}. Whereas the overmassive systems exhibit the typical unresolved dispersion spikes at their centres, this galaxy displays an anomalous central drop in its root-mean-square velocity ($V_{\rm rms}$) field, lacking the characteristic kinematic signature of a central dark mass.

This unexpected diversity suggests that while dissipationless mergers systematically move galaxies upward in $M_{\rm BH}$ at fixed $\sigma$, the growth of BHs at the extreme high-mass end is intrinsically stochastic. Such scatter can be naturally explained by extreme multi-body dynamics. For example, the coalescence of a supermassive BH binary generates anisotropic emission of gravitational radiation, imparting a net linear momentum recoil kick to the merged remnant \cite{Campanelli2007, Blecha2016, Komossa2012}. In the shallow, core-scoured stellar density profiles characteristic of ultramassive galaxies \cite{Faber1997,Kormendy2009}, dynamical friction is exceptionally weak. Even a sub-escape recoil kick can readily displace the BH from the galactic nucleus onto a wide, long-lived wandering orbit \cite{Gualandris2008}. Alternatively, successive dry mergers in dense environments may introduce a third supermassive BH before an existing binary has coalesced \cite{Merritt2005}. Chaotic three-body gravitational slingshot interactions typically eject the least massive object into intergalactic space \cite{Hoffman2007}, truncating the monolithic accumulation of central dark mass---a scenario recently supported by observations of a candidate runaway supermassive black hole \cite{vanDokkum2026}.

Either of these dynamical processes could explain why some ultramassive galaxies do not host their expected ultramassive BHs at their optical centers today. Larger, volume-complete statistical samples of ultramassive galaxies observed with the spatial resolution of JWST will be required to determine the absolute frequency of this phenomenon, and to definitively map the interplay between systematic dry-merger growth and stochastic multi-body ejections in the most extreme galaxies in the Universe.

\section*{Methods}
    
\subsection*{Observational Data and Dynamical Modelling}
Rather than condensing the extensive observational and modelling procedures into this section, we refer the reader to two dedicated companion papers for full technical details. In brief, Paper I \cite{Cappellari2026p1} outlines the parent selection of the JWST Ultramassive Galaxy Sample ($M_\star \gtrsim 2\times10^{12} M_\odot$). It details the derivation of the deprojected stellar mass distributions via Multi-Gaussian Expansion (MGE) modelling \cite{Cappellari2002mge} of high-resolution JWST/NIRCam \cite{Rieke2023} imaging combined with wide-field ground-based DESI Legacy Surveys data \cite{Dey2019}. It also presents the extraction of the stellar kinematics from JWST/NIRSpec integral-field spectroscopy \cite{Boker2022} using the penalized pixel-fitting (\textsc{pPXF}) method \cite{Cappellari2004,Cappellari2017,Cappellari2023}. The exceptional stability of JWST provides diffraction-limited integral-field kinematics with a target $S/N \approx 80-100$, easily resolving the BH sphere of influence. 

Paper II \cite{Cappellari2026p2} presents the subsequent dynamical modelling. It employs the novel spectral solver for Jeans Anisotropic Modelling (JAM) \cite{Cappellari2008, Cappellari2020, Cappellari2026jam}, natively handling general velocity anisotropy and marginalizing over the unknown intrinsic geometry of these systems (rigorously evaluating both oblate and prolate extremes). By fitting the high-fidelity photometric models and two-dimensional kinematics in a Bayesian framework, Paper II derives the highly robust central BH masses utilized in this present work.

To place these JWST targets on the BH scaling relations, we systematically derived their effective velocity dispersions ($\sigma_{\rm e}$) and total stellar masses ($M_\star$). The dispersion was measured using \textsc{pPXF} from the global co-added spectrum within the NIRSpec field of view (shown in fig.~7 of Paper I). This measured value was then aperture-corrected to one effective radius ($1R_{\rm e}$) using a standard empirical formulation \cite[eq.~1]{Cappellari2006} appropriate for massive early-type galaxies \cite{Zhu2023sigma}, where the circularized $R_{\rm e}$ was computed directly from the MGE models using the \texttt{jampy.mge.half\_light\_isophote} routine. Finally, the total stellar mass was computed by multiplying the total analytic luminosity of the MGE by the best-fitting stellar mass-to-light ratio ($M/L$) derived from the prolate JAM models. Because dark matter is expected to be dynamically negligible within the innermost galaxy centre \cite[fig.~11]{Cappellari2026}, this provides an exceptionally reliable measure of $M_\star$. Crucially, unlike mass estimates based purely on stellar population synthesis, this dynamical approach automatically accounts for any variations in the stellar initial mass function (IMF), which can alter the true $M/L$ at fixed stellar population parameters.

\subsection*{Compilation of Dynamical Black Hole Masses}
To place the eight JWST measurements in the context of the local supermassive black hole (SMBH) population, we constructed a comprehensive, fully auditable master compilation of spatially resolved dynamical BH mass measurements. This database serves as a superset of secure BH detections drawn from major historical compilations \cite{Gultekin2009, Beifiori2012, McConnell2013, Kormendy2013review, Saglia2016, vandenBosch2016, Sahu2019}, systematically augmented with the latest high-precision dynamical measurements available in the literature. These recent additions include advanced stellar-dynamical modelling utilizing both Jeans Anisotropic Modelling (JAM) and Schwarzschild orbit-superposition methods \cite[e.g.][]{McConnell2012, Liepold2023, Simon2024, Thater2026}, as well as spatially resolved CO gas-dynamical determinations from ALMA, including those produced by the WISDOM project \cite[e,g,][]{Davis2013, Onishi2017, Davis2017, Dominiak2025}. To ensure the highest accuracy for our scaling relations, we inspected the relevant papers and we strictly excluded upper limits (such as those in \cite{Beifiori2012}) and preliminary or uncertain estimates \cite[e.g.][]{Cappellari2008iaus,deNicola2026}.

To maximize reproducibility and preserve exact observational and modelling provenance, the master catalogue is structured as a \emph{measurement-level} database rather than a one-row-per-galaxy table. All independent BH determinations and varying dynamical assumptions applied to the same data (e.g., differing JAM alignment assumptions for NGC 4751) are retained as distinct entries, unified by a canonical galaxy identifier. Crucially, however, these multiple entries are not treated as independent data points in the statistical analysis. To evaluate astrophysical scaling relations---where the fundamental statistical unit must be the physical galaxy---we aggregated the measurement-level catalogue into a galaxy-level sample. For any galaxy possessing multiple retained $M_{\rm BH}$ determinations, we adopted the median of the available black hole masses as the representative physical value. 

\subsection*{Assessing Recent Ultramassive Black Holes}
This section evaluates recent studies of a sample of 16 brightest cluster galaxies \citep{deNicola2025, deNicola2026}. \citet{deNicola2025} establish BH scaling relations for the sample, while \citet{deNicola2026} present the underlying dynamical models and kinematics. However, detailed models are only published for one galaxy: the system with the largest BH mass. Interestingly, rather than displaying a typical central peak in velocity dispersion, this galaxy shows a central drop.

The unpublished figures of the triaxial Schwarzschild model fits for the 16 galaxies, the raw spectra at the various slit positions, the extracted kinematics, and the photometric profiles were provided by private communication from the lead author. A detailed inspection of these data revealed the fundamental limitations of ground-based observations for these specific targets. The seeing-limited, long-slit spectra suffer from poor spatial resolution and low signal-to-noise ratios, with a median seeing FWHM of 1.0 arcsec and a median $S/N \approx 22$ across all galaxies and observed position angles (PAs). We robustly computed this $S/N$ directly from the residuals of our \textsc{pPXF} spectral fits, matching our methodology for the NIRSpec data in Paper~II. These ground-based limitations stand in stark contrast to our diffraction-limited JWST IFS data, which achieves a sharp spatial resolution of 0.15 arcsec FWHM and an $S/N \approx 70$.

In principle, the line-of-sight velocity distribution (LOSVD) can be extracted at any $S/N$ level, with lower data quality ostensibly just yielding larger, yet still meaningful, statistical uncertainties. However, this ignores the critical interplay between random noise and systematic effects. While systematics are negligible at high $S/N$, at low $S/N$ levels they can easily overwhelm the weak signal, severely biasing the extraction away from the true solution and rendering formal statistical errors meaningless. Consequently, extracting the full shape of the LOSVD---which is strictly required to reliably constrain Schwarzschild models beyond just $V$ and $\sigma$---becomes highly degenerate and unreliable. To ensure robust extraction free from such biases, previous studies consistently recommend $S/N \gtrsim 50-60$ per resolution element \cite{Bender1994, Cappellari2004}. Crucially, this includes thresholds of $S/N \gtrsim 50$ recommended specifically for the kinematic extraction techniques \cite{FalconBarroso2021} employed on these galaxies by \cite{deNicola2026}. Falling well below these thresholds, the provided kinematic profiles were highly noisy, contained numerous masked data points, and exhibited large, unexplained inconsistencies across different position angles.

Only two out of the 16 galaxies showed clear evidence of an increase in $\sigma$ towards the galaxy centre, as is dynamically expected for ultramassive BHs (and as unambiguously observed in 7 of the 8 JWST targets). For the remainder of the sample, it was impossible to infer whether the observed variations in the nuclear $\sigma$ profiles were a confirmation of the stochasticity we unambiguously discovered in our JWST sample, or simply noise and systematic artefacts.

To quantitatively assess the robustness of previous claims \cite{deNicola2026}, we constructed MGE models from the provided photometric profiles and fitted JAM models to the extracted kinematics, as we did for our JWST data. Except for the two galaxies with a clear central rise in $\sigma$, the severe noise in the data made it impossible to fit convincing BH masses. The BH masses formally derived for the remaining galaxies were generally $<10^{10} M_\odot$ and showed virtually no correlation with previously reported values \cite{deNicola2025}. This stands in stark contrast to the strong agreement found between JAM and Schwarzschild BH determinations in 33 other systems, a comparison sample that notably included several triaxial galaxies \cite[fig.~1]{Cappellari2026p2}. We conclude that the 16 BH masses reported for this sample \cite{deNicola2026} are not robustly constrained by the long-slit data used to measure them. For this reason, these measurements have not been incorporated into the analysis in this paper.

\subsection*{Scaling Relations Analysis}
Standard robust linear regressions utilizing a $2.6\sigma$ clipping threshold were employed to derive the quoted best-fit scaling relation slopes and intrinsic scatters. We fit the relations using the \texttt{ltsfit} package \cite{Cappellari2013p15}, which accounts for the intrinsic scatter in the relation, reliably identifies true astrophysical outliers while isolating the primary sequence of the broad comparison sample. 

\subsection*{Data Availability}
All observational data, extracted kinematics, and derived data products underlying this article will be made available with the published paper. 

\subsection*{Code Availability}
The software packages used for the dynamical modeling and statistical analysis in this work are open-source and publicly available via the Python Package Index (PyPI). The spatial binning package \textsc{PowerBin} of \cite{Cappellari2025} is available at \url{https://pypi.org/project/powerbin/}. The spectral Jeans Anisotropic Modelling (JAM) software \cite{Cappellari2026jam} is available at \url{https://pypi.org/project/jampy/}. The penalized pixel-fitting (\textsc{pPXF}) software \cite{Cappellari2023} is available at \url{https://pypi.org/project/ppxf/}. The Multi-Gaussian Expansion (MGE) software \cite{Cappellari2002mge} is available at \url{https://pypi.org/project/mgefit/}. The robust linear regression package \texttt{ltsfit} \cite{Cappellari2013p15} is available at \url{https://pypi.org/project/ltsfit/}.

\backmatter 

\bmhead{Acknowledgements}

This work is based on observations made with the NASA/ESA/CSA James Webb Space Telescope. The data were obtained from the Mikulski Archive for Space Telescopes at the Space Telescope Science Institute, which is operated by the Association of Universities for Research in Astronomy, Inc., under NASA contract NAS 5-03127 for JWST. These observations are associated with JWST programme 8217. 

The authors used generative AI assistants to improve workflow efficiency, including language editing and programming support; all scientific ideas, analyses, interpretations, and conclusions are their own, and the authors take full responsibility for the content of this work.

\bmhead{Author contributions} All authors contributed extensively to the work presented in this
paper.

\bmhead{Competing interests} The authors declare no competing interests.

\end{document}